\documentstyle[prl,aps]{revtex}
\vspace{-8pt}
\title{Nano-Ploughed Josephson Junctions\\
as on-chip Radiation Sources.}

\author{B.~Irmer, F.~Simmel, R.H.~Blick, H.~Lorenz and J.P.~Kotthaus \\ 
{\normalsize\it Center for NanoScience and Sektion Physik,
Ludwig-Maximilians-Universit\"at M\"unchen, Germany.} }
\author{M.~Bichler and W.~Wegscheider
\\ {\normalsize\it Walter Schottky Institut, TU M\"unchen,Germany.}}

\def\LaTeX{L\kern-.25em\raise.425ex\hbox{a}\kern-.075em\TeX}

\def\fakebold#1{\relax\ifvmode\leavevmode\fi%
\ifmmode%
\setbox0=\hbox{$#1$}%
\else%
\setbox0=\hbox{#1}%
\fi%
\kern-.02em\copy0 \kern-\wd0%
\kern .04em\copy0 \kern-\wd0%
\kern-.0125em\raise.02em\box0%
}%
\begin{document}
\label{firstpage}
\maketitle
\sloppy
%\begin{center}
%\received{(Received today)}
%\end{center}
%**********************************************************************%
%
\begin{abstract}
A new technique is presented which enables the fabrication of highly
transparent Josephson junctions in combination with mesoscopic devices.
We utilize a modified AFM tip to plough grooves into superconducting material,
thus defining a weak link. This weak link is made within the superconducting
split-gates, which are used to electrostatically form a conventional quantum dot and serves as a source
of millimeter wave radiation around 100~GHz. We show the characteristics
of typical junctions built and discuss their high-frequency properties.
We find that the millimeter wave emission of the weak link leads to a bolometric
effect in the case of quantum point contact spectroscopy.
\end{abstract}

\section{Introduction}
Direct mechanical structuring seems not to be the obvious method to build mesoscopic devices.
Nevertheless, we applied the fairly ancient technique of ploughing~\cite{stone} with
modified AFM-tips to build
Josephson junctions (JJs)~\cite{wir} in conjunction with standard mesoscopic systems,
e.g. quantum point contacts~\cite{qpc88}, single and double quantum dots~\cite{kouwenhoven97}. 
The main advantages of
our approach are the ease of integration of the different techniques and the flexibility to
build spectroscopic tools in the quantum limit.
While conventional spectroscopy on mesoscopic structures relies on bulky external frequency
sources, our aim is to place the millimeter wave sources in nanometer distance to the
device under test. This not only allows the generation of high intensity electromagnetic
fields exactly at the nano-structure, minimizing effects by external heating, but also
spectroscopy on quantum structures and the investigation of coherent electron tunneling
phenomena directly in the frequency domain.\\
In the first part of this work we focus on the general properties of JJs, on 
the fabrication and present characteristic measurements
of these millimeter wave
sources. Moreover, we discuss the suitability of the ploughing technique to mechanically
form coupled JJs. 
%In the following paragraph we will discuss the integration into a common
%double quantum dot device. 
Then we present first measurements on
a quantum point contact coupled to a single junction's radiation and finally we discuss
the possibility of applying the sources
for spectroscopy on quantum dot devices.
%That´s your job, Robert...\\

%%antenna, cavities\\ solution: on-chip radiation sources\\

%how to make? JJ!

% \subsection{Decoupling Radiation from Josephson Junctions.}
%A voltage across the junction increases the phase difference with
%time, which then precesses as $\partial\Phi/\partial
%t=2eV/\hbar=\omega_J$. The efficiency with which radiation can
%decouple depends not only on the material specific permittivity
%$\epsilon_r$ but also on the junction geometry.

%The propagation speed becomes
%$\tilde{c}=c_0/\sqrt{\epsilon(1+2\lambda_L/L)}$~\cite{swihart}.
%The characteristic impedance is then
%$Z_J=L/W\cdot\sqrt{(1+2\lambda_L/l)/\epsilon}\cdot
%Z_0$~\cite{swihart} with $Z_0$=377~$\Omega$. In oxide junctions
%the length $L$, whereby $L \ll \lambda_L$, is defined by the
%thickness of the tunneling barrier. This makes $\tilde{c} \ll c_0$
%and the coupling to the environment very inefficient.

\section{Fabrication}

\subsection{Why ploughing the Josephson-Junctions ?}
In the history of prediction and discovery of the Josephson
effect~\cite{joseph} a major role was played by tunnel
junctions, where the two superconducting reservoirs are weakly
connected by an thin insulating barrier. However, the coupling
between the reservoirs can be realized by any means as long as it
is ``weak''´ - i.e. the order parameter is suppressed locally but
the reservoirs become not completely independent from each other.
In very different types of junctions the Josephson effect has been
observed, e.g. point contacts \cite{zimmerman66}, proximity effect bridges \cite{notaris},
superconductor-normal-superconductor systems and micro patterned
thin film systems (Dayem bridge~\cite{dayem1} and Variable Thickness
Bridge~\cite{vtb}) to name only the most prominent.\\ 
The radiation generated by the ac Josephson current should couple
into the environment with a non vanishing efficiency and the
fabrication of the junction has to be fully compatible with the
preparation of the quantum dot system. As an example
Fig.~\ref{dd+wl_sem} shows an electron microscope image of a
GaAs/AlGaAs heterostructure with Aluminium gates on top of it,
which electrostatically define two quantum dots separated by
tunable tunnel barriers. The JJ has to be placed
as close as possible to the quantum dots (precision in alignment)
and its integration has to be fully compatible with the quantum
dot system (best made out of the same thin film). This adds
serious technological complications.\\

In the following we
will discuss their use for this purpose from a technological point
of view as well as from their ac-Josephson behaviour~:\\
Tunneling JJs require high quality and very thin (thickness $d\approx$
10 to 20~\AA) oxide barriers. This is commonly achieved by thermal
oxidation of the first electrode, on top of which the second
metallic electrode is placed. This sets very high tasks into the
patterning and alignment and minimizes the yield in this process
when combined with the quantum dot definition, which is non-trivial itself.
Furthermore, tunneling
junctions suffer from comparably low critical Josephson current
densities $j_c\approx10^{2}$A/cm$^2$~\cite{Likharev}, which are
limited by the tunnel resistance $R_n\approx 1~$k$\Omega$.
The efficiency with which radiation can
decouple depends not only on the material specific permittivity
$\epsilon_r$ but also on the junction's dimension:
the wave propagates with the Swihart velocity $\tilde{c}=c_0/\sqrt{\epsilon_r(1+2\lambda_L/L)}$~\cite{swihart}, wherein 
$\lambda_L$ is the London penetration depth and $L$ the length of the barrier. 
The impedance of the JJ is then
$Z_J=c_0/\tilde{c} \cdot (L/\epsilon_r W) \cdot
Z_0$~\cite{swihart} with the free space impedance $Z_0$=377~$\Omega$ and $W$ the width of the barrier. In oxide junctions
the length $L$ is defined by the
thickness of the tunneling barrier, whereby $L \ll \lambda_L$. This results in $\tilde{c} \ll c_0$
and the coupling to the environment will be very inefficient, e.g. for $L\approx$10~\AA the impedance becomes $Z_J\approx 10^{-4}Z_0$.
Another
important aspect is the large intrinsic capacitance $C$ of the
junction, which effectively shunts the tunnel conductance at
higher frequencies~\cite{Likharev78}.
\\

Microbridges, however, are in general expected to show nearly
ideal Josephson behaviour if its dimensions $L_{eff}$ are
sufficiently small relative to the coherence length
$\xi$~\cite{klapwijk}. Their main advantage is the lower
resistance $R_n\approx 0.1~\Omega$, and hence the higher critical
current densities $j_c$ of typically $10^{6}$A/cm$^2$. This is
expected to result in a dramatically enhanced microwave emission.
%power $P_J=R_n \cdot I_c^2$.
Furthermore the length $L$ of the
bridge can be made comparable to the $\lambda_L$,
which results in a better impedance matching between the
junction and the environment, e.g. for $L=100~$nm $Z_J$ becomes $1/3~Z_0$.\\
%Microbridges can in principle be made from and in
%fact into the same thin film already defining the gates of the
%quantum dot structure.
%\\
Common techniques for weak link fabrication are cutting thin films
with razor blades or diamond
scratchers~\cite{dayem},\cite{feldman},\cite{mooij}. Although
bridges down to some 0.1 $\mu$m have been made, they can not be
positioned into predefined gate structures with the same
precision. 
%The same problems apply for e-beam lithography and
%lift-off technique.
Nevertheless, nanometer precision is standard for
scanning probe microscopes, both in imaging and
patterning~\cite{wendel}. By giving the tip the functionality of a
tool, e.g. a plough, one can accurately position the tool
within the gate structure of the quantum dot device, remove thin
film material in a well defined way, leaving behind trenches which
then define microbridges of constant or variable
thicknesses. The advantages of applying a nano-plough for
lithography are obviously the precision of alignment, the
non-damaging definition process compared to electron or ion beam
lithography, and the absence of additional processing
steps, such as etching the substrate.\\ We define our devices in
Al thin films with a thickness around 100~nm, which are
thermally evaporated\footnote{In order to achieve reasonable
superconducting properties, we use high evaporation rates
$\approx$~100~\AA~and low background pressure,
typically at $p\le 10^{-7}$~mbar. The critical temperature $T_c$ is typically
slightly above 1~K for thin films.} onto semi-insulating GaAs substrates for the
$IV$-characteristics and GaAs/AlGaAs heterostructures for the
quantum point contact measurements. Wire and gate structures 
are predefined with optical and electron beam lithography.

\subsection{Tools used:~Nano-Ploughs}

A crucial point for nano-ploughing is the appropriate plough: common
scanning probe microscope
tips are either robust but too blunt to form small trenches
(Si$_3$N$_4$ or hard coated tips)
or too brittle for this purpose (single crystal Si). We employ {\sl
electron
beam deposition} (EBD) to grow material onto the end of common
Si tips
by focusing the beam of a scanning electron microscope.
The highly energetic electrons interact with additionally
introduced organic gases and form an organic compound, known as high
dense carbon~\cite{wendel}. This material shows a hardness comparable to that of
diamond~\cite{bert} in combination with being non-brittle~\cite{nanotools}.
Special care has to be taken to
ensure a robust interface between the EBD-tip and the silicon pyramid. This is done by growing first a tripod onto the sides of
the Si tip before growing the plough itself. This gives high stability to withstand shear forces.
For the nano-plough 500~nm long and 50~nm wide needle-like tips are
grown under a non-zero angle
with respect to the axis of the pyramidal Si tip.
When dragged through
material this angle causes an additional vertical force component
which ensures the plough to be pushed downwards, cutting its way through the
material.\\

\subsection{Ploughing of superconducting thin films}
Imaging and positioning is performed in standard
AFM-mode~\cite{tapping}. In the ploughing-mode, the tip is
displaced by nominally $\Delta z$ towards the surface, resulting
in a loading force $F_{\perp} \approx k \cdot \Delta z$.
Cantilevers with high force constants $k \approx 40 - 100~$N/m are
used. The vertical displacement ranges from $\Delta z \approx 0.1~\mu$m
to several micrometers, offering a wide range of loading forces.
These can be sufficient to completely remove the metal layer or to
form a trench with residual base thickness. To define a
microbridge of width $W$, the tip is withdrawn from the surface,
displaced by a length $W$ and driven into the metal layer again. The
length $L$ of the Josephson contact is defined by the width of the
trench, which again is given by the tip diameter. To date we were
able to cut 300~nm thick Al films with a minimum line width of
50~nm, yielding an aspect ratio of 1~:~6. In
Fig.~\ref{paralleleJJ} several trenches under various ploughing
conditions are shown.
For all lines the same tip has been used and the ploughing speed
was 100 nm/s throughout. The vertical displacement $\Delta z$
is increased from $0.5 ~\mu$m for the top line to $2.5 ~\mu$m at
the base line. For the indentation pressure involved we can
estimate 1 to 100 GPa, whereas the hardness of evaporated Al is
about 0.5 GPa~\cite{hay98}. The Al film is cut in a very
controlled way, the removed material is placed on both sides
of the trench. In this way multiple JJ arrays can
be made with a spacing close to $L$. Weak scratches in
the GaAs substrate can be seen, thus ensuring that all material
has been removed from the film.\\
With appropriate
parameters, either Dayem bridge type weak links with constant film
thickness, or variable thickness bridges or even a combination of
both can be fabricated. As an example a Dayem bridge with
dimensions $W \times L = (100 \times 100)~$nm$^2$ is shown in the left
part of Fig.~\ref{paralleleJJ}.

\section{Characterisation of Josephson Junctions}

\subsection{High transparency}

All transport measurements are conducted in a standard current feed four-probe
setup in a dilution refrigerator at 35~mK.
%Microwave radiation
%of $f=20 Ghz$ is   no cavity and no antenna used.
First we look at the temperature dependence of the critical
current density (Fig.~\ref{KO}), which is estimated to be $j_c(T\rightarrow 0) =
I_c / L\cdot d = 5.1 \times 10^{6}~$A/cm$^2$, where $d$ is the thickness of the film. From the temperature
dependence of the critical voltage $V_{c}=I_{c}\cdot R_{n}$
(Fig.~3) we reproduce the universal, material independent slope
$\alpha = \partial(R_N \cdot I_c) / \partial T = 2\pi k_B /
7\zeta(3)e$ for $T \rightarrow T_{c}$. This also identifies the
conduction mechanism inside the junction: the
Kulik-Omelyanchuk-theory (KO: straight line)~\cite{kulik} for the
case of a clean weak link, i.e. $\xi, l \gg L_{eff}$, where $l$ is
the mean free path of electrons and $\xi$ the coherence length, fits very well to our data. A
tunnel junction in comparison gives much lower critical
currents (Ambegaokar-Baratoff: AB) \cite{ambegaokar}.
From its critical temperature we calculate the BCS gap
$\Delta_{Al}(T\rightarrow 0) = 3.37 / 2\cdot k_B T_c = 138~\mu$eV
\cite{BCS}. The associated clean limit coherence lenght $\xi_{Al}
= 0.18\cdot \hbar v_F / (k_B T_c)$ is then $\xi_{Al} =
2.9~\mu$m~\cite{tinkham}, using the bulk Fermi velocity of $v_F =
2.03 \times 10^6~$m/sec. This should be compared to the effective
length $L_{eff}$ of the Josephson constriction, which is somewhat
larger than the 100~nm geometric length, but still an order of
magnitude smaller than $\xi$, which makes our junction extremely
transparent~\cite{Likharev}\cite{zaitsev}.

\subsection{Dayem bridge}
Here we only want to describe the main features in the
current-voltage characteristic for the two types of microbridges
and especially their differences. A detailed description of the
observed phenomena is e.g. given in \cite{barone}.\\
In Fig.~\ref{IV-dayem} a set of
characteristic $IV$-trace of a single Al junction is shown,
where the external magnetic field is increased from zero to 10
mT. The dc-Josephson current can be clearly seen. For currents
larger than the critical current $I_c$, a finite voltage drops
across the junction, defined by its normal resistance $R_n =
0.57~\Omega$. The excess current is close to zero. The
$IV$-characteristic is not hysteretic and single valued. This
agrees well with the theoretical description for a non-tunneling,
Dayem-bridge like superconducting weak link described by
Likharev~\cite{Likharev}. For increasing magnetic fields the
Josephson current is reduced to zero, i.e. normal
conductivity, at $B_c=18~$mT.

\subsection{Variable Thickness Bridge (VTB)}

For the VTBs a very distinct behaviour is observed: starting from $V=0$ a
supercurrent accompanied by a finite voltage is observed up to about
115 $\mu$V, characterised by a differential resistance
$R_d=24.7~$m$\Omega \ll R_n=0.48~\Omega$. The $IV$-charateristic
progresses, after a small jump in voltage, into a curve with higher
resistance. This section shows a significant fine-structure. At
even larger voltages the $IV$ curve bends towards the normal
current carrying state defined by $R_n$, probably due to heating.
At this point the extrapolated characteristic goes exactly through
$I=V=0$.
%The absolute value of $I_c$ is almost 10 times larger
%than that observed in the dayem bridge.
This behaviour is very similar to that observed by Klapwijk {\sl
et al.}~\cite{klapwijk76}, where it is interpreted as a
consequence of a phase slip process. A flux flow hypothesis is
inapplicable because the sample size is too small to allow vortex
penetration~\cite{barone}. The fine structure can be related to
self resonant modes, where the electromagnetic field associated
with the ac-current interacts with the external environment and
the junction cavity.

\subsection{HF-response}

In order to verify the ac Josephson effect and the microwave
response of the nano-ploughed bridges, radiation of $f=20$~GHz
from an external source is radiated onto the microbridge. No on chip
antenna or special cavity were used. Clearly the junction does
interact with the electromagnetic field and changes its  $IV$
characteristic significantly. The peculiar step-like shape of it
can be understood as a superposition of externally induced Shapiro 
steps and self resonant modes of the intrinsic Josephson
radiation~\cite{PRL}.

\section{Josephson junction coupled to a quantum point contact}

%\subsection{Fabrication of the Josephson junction spectrometer}
\subsection{Bolometric response of a quantum point contact to Josephson radiation}

The nano-ploughed Josephson junctions can be easily combined with
mesoscopic semiconductor structures such as quantum point contacts or quantum dots 
(Fig.~\ref{dd+wl_sem}). One of the gates is equipped with four terminals to ensure full control of the JJ. 
The junction itself is defined by nano-ploughing at the desired position (as described in Section 2).
The two-dimensional electron gas (2DEG) of the GaAs/AlGaAs heterostructure we use
is located 50~nm below the surface. It has a sheet density of $n_S=1.6 \times 10^{11} $cm$^{-2}$ and a 
low-temperature mobility of  $\mu=8 \times 10^5 $cm$^2/$Vs.\\
As a first demonstration of the influence of the microwave radiation emitted by the weak link
on a mesoscopic structure in the 2DEG we report on the bolometric change in the quantized conductance steps
of a quantum point contact (QPC) \cite{qpc88}.
Microwave radiation causes heating of the electron reservoirs separated by the QPC which leads to
a bolometric photo-conductance signal.
The current through a QPC can be written as
\[
I=\frac{2e}{h}\int_0^{\infty} t(E) [f(E-\mu_S;T_S)-f(E-\mu_D;T_D] dE,
\]
where $t(E)$ is given by $t(E)=\sum_n \theta(E-E_n)$ with $E_n$ being the $n$th subband energy of the QPC,
and $\mu_S,T_S$ and $\mu_D,T_D$ the chemical potentials and the local temperatures of source and drain,
respectively \cite{wyss}.
The chemical potentials differ by the bias voltage $V_{sd}$  (cf. Fig. \ref{figure7}).
The difference between the Fermi
distributions of the electrons on either side leads to an enhancement of the current
in the regions between the quantized conductance steps \cite{wyss}.\\
In the experiment one of the quantum point contacts of the quantum dot structure in
Fig. \ref{dd+wl_sem} was defined in the 2DEG via field-effect by applying negative voltages $V_g$
to the gates indicated.
Across the QPC a bias voltage of  $V_{sd}=300~\mu$V was applied.
The differential conductance of the QPC was measured at a  temperature of 35~mK using standard lock-in techniques.
The weak link was tuned into the ac Josephson regime  with a dc voltage of  $V_{JJ}=500~\mu$V
at zero magnetic field. The measured conductances and the difference signal
(photo-conductance) are displayed in Fig. \ref{figure9}. Switching on the weak link clearly results
in an enhancement of the conductance in the regions between the conductance plateaus as
expected from theory.
This demonstrates the feasibility of combining a weak link as an on-chip microwave source
with a mesoscopic structure in the plane of the 2DEG of a GaAs/AlGaAs heterostructure $50$~nm below.

\section{Applications in photon-assisted mesoscopic transport}

Apart from bolometric signals the phenomena one seeks to observe in the transport properties of
mesoscopic structures under the influence of microwave radiation are photon-assisted tunneling and Rabi oscillations.
In the tunneling regime, absorption or emission of a photon by a tunneling electron can enable tunneling processes
at energies $\pm n \hbar \omega$ normally forbidden by energy conservation
\cite{tien63,buettiker82}. These photon-assisted processes can significantly alter the transport properties
of quantum dots in the Coulomb blockade regime. Quantum dots are small electronic islands isolated
from the 2DEG by tunneling barriers \cite{kouwenhoven97}.
In the Coulomb blockade regime $k_BT \ll e^2/C$, where
$C$ is the total capacitance of the quantum dot. We find for typical devices, $T \approx 100$mK and $C \approx 100$ aF.
In this regime, transport through the quantum dot is normally blocked due to the large
charging energy for a single electron. However, at certain gate voltages the energy of the dot with $N$ electrons
is degenerate with the energy of the dot occupied by $(N+1)$ electrons. Then, the charge on the dot is allowed to
fluctuate resulting in a sharp conductance peak due to tunneling of single electrons onto and off the dot.
Photon-assisted tunneling under microwave irradiation
leads to the occurence of photon-sidebands of these conductance peaks \cite{kouwenhoven94,blick95,oosterkamp97}.\\ \\
Transport through a double quantum dot system is only possible if, in addition to the requirements
for a single quantum dot, an occupied electronic state in one dot is aligned with an accessible state in the other
dot. The resulting conductance peaks are even less broadened than in the single quantum dot case, as the thermal broadening
induced by the reservoirs is circumvented by this inter-dot resonance criterion.
Due to the diminished broadening of the conductance peaks photon-sidebands can be observed more clearly and at frequencies below the 
limit of thermal broadening (usually down to 1 GHz).
By appropriately tuning the bias and gate voltages a variety of transport scenarios can be realized
\cite{oosterkamp98}. One prominent example is the single electron pump \cite{stafford96}.
In a double quantum dot with weak inter-dot coupling, i.e. a large tunneling barrier and a small capacitance
between the two dots, electrons can be sequentially pumped through
the system under the influence of microwave irradiation as depicted in Fig. \ref{figure10}.
If, in contrast, the two quantum dots are strongly coupled, an artificial molecule will be formed.
Analogous to molecular physics the energy levels of the double quantum dot are then tunnel-split into
bonding and anti-bonding states. Irradiation with microwaves at a frequency corresponding to the tunnel-splitting
leads to Rabi oscillations within the artificial molecule \cite{blick98}.\\ \\
Observation of such photon-assisted phenomena is experimentally highly challenging
as these experiments require the combination of very low temperatures and microwave radiation.
Transport spectroscopy as described above has to be performed in a dilution refrigerator at temperatures of
$\approx 100$ mK.
This complicates the application of externally generated microwave radiation over a wide frequency range as one is
limited by losses in cables, heating effects and the space available in the refrigerator.
The coupling of  the radiation to the device under investigation has to be achieved with specially designed antennas.
Utilizing the ac Josephson effect for tunable on-chip microwave sources represents a promising alternative approach
to the realization of photon-assisted transport experiments circumventing the problems described.
The most significant difference to
usual experimental setups is the proximity of the microwave source to the structure of interest.
So far, it is not clear whether and how the near-field properties of the source modifies
photon-assisted transport.\\ \\
Another fascinating aspect of future experiments on a double quantum dot with an embedded weak link is its relation to
electrical quantum standards. Frequency locked pumping of electrons through multiple tunnel junctions has
been demonstrated to serve as a current standard yielding a current $I = e f$ \cite{geerligs90}
or as a single electron counter \cite{keller96}. In such devices
single electrons are pumped by appropriately raising and lowering tunnel barriers, i. e. gate voltages.
Due to the probabilistic nature of tunneling the accuracy of the pump decays rapidly for
frequencies higher than typically 10-50 MHz as then some electrons will not tunnel each cycle.
Furthermore the metallic gates are effectively shunted at high frequencies disabling controlled
turnstile operation. Thus the attainable pump current through such a turnstile is only a few pico-ampere.
It is an interesting question whether coherent tunneling through double dot systems
at GHz frequencies could pump accordingly higher currents. A combination of a double quantum dot and one or
several weak links enables us to investigate such pumping processes driven by the ac Josephson effect.\\

\section*{Acknowledgments}
We would like to thank  
W.~Zwerger, T.M.~Klapwijk and M.~B\"{u}ttiker for stimulating discussions,
A.~Kriele, S.~Manus and W.~G\"odel for their continuous support.  
We also thank R.J. Warburton for a critical reading of 
the manuscript.
The work was funded by the Volkswagen foundation under grant \# 
I/68769 and the Deutsche Forschungsgemeinschaft (SFB 348).

% de oide Schwungscheibn.

\begin{figure} [p]
   \caption{\rm SEM-micrograph of a double quantum dot with an embedded nano-ploughed weak-link. The white
    arrows indicate the gates swept for the QPC measurement shown in Fig.\ 9. Circle indicates the position of the JJ.}
\label{dd+wl_sem}
\end{figure}

\begin{figure}
    \caption{\rm Left: a schematic sketch of the nano-ploughs used. Note the side coverage to ensure a stable and robust interface between
    the basic silicon tip and the EBD-plough. Right: Electron microscope image of a typical tip.}
\label{tips}
\end{figure}

\begin{figure}
    \caption{\rm Nano-ploughed lines in an Al-film. 
	Left: a typical Dayem style microbridge with dimension $L \times W=100~$nm$ \times 100~$nm.
	Right: the loading force is increased
    from top to bottom, defining variable thickness bridges and bridges of increasing length $L$. The
    junction width $W$ is easily adjusted by a horizontal displacement $W$, where the plough is withdrawn from the film.
    }
\label{paralleleJJ}
\end{figure}

\begin{figure}
   \caption{\rm Left: from the temperature dependence, our junction (black dots) is interpreted as a
                clean weak link with $L \ll l, \xi$ (Kulik and Omelyanchuk: KO). KO theory predicts a
critical current of $\pi$ for $T/T_{c}=0$.
The universal slope $\alpha = \partial(R_N \cdot I_c) / \partial T$
for
$T \rightarrow T_{c}$ implies $\alpha = 615 \mu$V/K in
good agreement with the theoretical value $\alpha = 635 \mu$V/K. This
device shows a critical temperature T$_{c}$= 950~mK. Right: characteristic IV trace of a dayem bridge style junction at T=35~mK.}
\label{KO}
\end{figure}

\begin{figure}
   \caption{\rm IV-characteristic of a Dayem bridge style Josephson junction under various magnetic 
   fields perpendicular to film.
   The critical field $B_c$ is 18 mT. The inset show a schematic view and a micrograph of the actual device.}
\label{IV-dayem}
\end{figure}

\begin{figure}
    \caption{\rm $IV$-characteristic of a variable thickness bridge under various magnetic fields.
    The critical field $B_c$ is 20 mT. The inset show a schematic view and a micrograph of the actual device.
    Note that the current and voltage axis are enlarged by a factor 10 in comparison to Fig.~\ref{IV-dayem}.}
\label{IV-VTB}
\end{figure}

\begin{figure}
  \caption{\rm Left: $IV$-characteristic of a Al variable thickness microbridge without (0~GHz) and under
    external microwave radiation of $f=20~$GHz.
     Right: detailed view on the steps appearing.
    }
\label{hf}
\end{figure}

\begin{figure}
   \caption{\rm Different local temperatures of source and drain are the origin of the bolometric response of
a QPC to microwave irradiation.} 
\label{figure7}
\end{figure}

\begin{figure}
\caption{\rm Measurement of the bolometric response of a QPC to Josephson radiation emitted by
    the nano-ploughed weak link. The step-like curves show the conductance through the QPC without
	(solid line) and with radiation (dotted line). Substracting the signals with and without radiation, we obtain the 
	bolometric signal displayed to the right axis.}
\label{figure9}
\end{figure}

\begin{figure}
    \caption{\rm Left: Schematic representation of the double quantum dot with
	weak link. Right: Electron pumping through a double quantum dot under the influence of microwave radiation.
	Transport through the system is made possible by absorption of photons.}
\label{figure10}
\end{figure}

\label{lastpage}


\begin{thebibliography}{0}
\bibitem{stone}
The Sumerians not only invented the wheel around 3700 BC, but also the plough 3200 BC which
both can be regarded as major milestones in the history of human civilisation.
\bibitem{tapping}Imaging is done in dynamic AFM, commonly called
Tapping-Mode. This ensures a nondestructive imaging even when
cantilevers with high force constants are used.
\end{thebibliography}
\end{document}